\documentclass[12pt]{iopart}

\usepackage{bm,color}
\begin{document}

\title[]{A parametrization of two-dimensional turbulence based on a maximum entropy production principle  with a local conservation of energy}

\author{Pierre-Henri Chavanis$^1$\footnote{Corresponding
author: chavanis@irsamc.ups-tlse.fr}}

\address{$^1$ Laboratoire de Physique Th\'eorique, Universit\'e Paul
Sabatier, 118 route de Narbonne 31062 Toulouse, France}

\ead{chavanis@irsamc.ups-tlse.fr}

\begin{abstract}
In the context of two-dimensional (2D) turbulence, we apply the maximum entropy
production principle (MEPP) by enforcing a local conservation of energy. This
leads to an equation for the vorticity distribution  that conserves all the
Casimirs, the energy, and that increases monotonically the mixing entropy
($H$-theorem). Furthermore, the equation for the coarse-grained vorticity
dissipates monotonically all the generalized enstrophies. These equations may
provide a parametrization of 2D turbulence.  They do not generally relax towards
the maximum entropy state. The vorticity current vanishes for any steady state
of the 2D Euler equation. Interestingly, the equation for the coarse-grained
vorticity obtained from the MEPP turns out to coincide, after
some algebraic manipulations, with the one obtained with the anticipated
vorticity method. This shows a connection between these two approaches when the
conservation of energy is treated locally. Furthermore, the
newly derived  equation, which incorporates a diffusion term and a drift term,
has a nice physical interpretation in terms of a selective decay principle. This
gives a new light to both the MEPP and  the anticipated vorticity method.
\end{abstract}

\vspace{2pc}

\maketitle

\section{\label{intro}Introduction}

A remarkable property of two-dimensional (2D) turbulent flows is their
ability to organize spontaneously into coherent structures such as
large-scale vortices and jets (Bouchet and Venaille 2012). A famous
example of this self-organization is Jupiter's Great Red Spot (GRS), a
huge anticyclonic vortex persisting for more than three centuries in
the southern hemisphere of the planet. This self-organization of 2D
turbulence into large-scale vortices shares fascinating analogies with
the self-organization of stellar systems in astrophysics (Chavanis {\it et
al} 1996, Chavanis 2002).

Basically, geophysical and astrophysical flows are described by the 2D
Euler equations or by their generalizations (quasi-geostrophic
equations, shallow-water equations, primitive equations...). The 2D
Euler-Poisson system is known to develop a complicated mixing process
generating vorticity filaments at smaller and smaller scales. At the
same time, this mixing process leads to the formation of coherent
structures at large scales which look quasistationary provided that a
coarse-graining procedure is introduced to smooth out the filaments.
In order to explain this self-organization, a statistical theory of
the 2D Euler equation has been proposed by Miller (1990) and Robert
and Sommeria (1991). This statistical theory is the counterpart of the
theory of violent relaxation proposed by Lynden-Bell (1967) for the
Vlasov-Poisson system describing collisionless stellar systems such as
elliptical galaxies. In the Miller-Robert-Sommeria (MRS) theory, the statistical equilibrium state
of the 2D Euler equation (most probable or most mixed
state) is obtained by maximizing a mixing entropy while conserving
the energy and the infinite family of Casimirs.

Of course, the evolution of the coarse-grained vorticity
$\overline{\omega}({\bf r},t)$ which averages over the filaments and
which relaxes towards a quasistationary state is {\it not} given by
the 2D Euler equation.  We expect that it satisfies a kinetic equation
that relaxes towards the maximum entropy state. Actually, the
relaxation towards the maximum entropy state is not granted since it
depends on an hypothesis of ergodicity (or at least efficient mixing)
that is not always satisfied. Indeed, there are many situations in
which the evolution of the system is non-ergodic so that the QSS
differs from the statistical prediction.

An interesting problem is to determine the kinetic equation satisfied
by the coarse-grained vorticity $\overline{\omega}({\bf r},t)$. This
is interesting not only at a conceptual level, but also at a practical
level. Indeed, it is generally not possible to solve the 2D Euler
equations for the fine-grained vorticity ${\omega}({\bf r},t)$ exactly
because they develop small-scale filaments that ultimately lead to
numerical instabilities. In addition, we are generally not interested
in the small scales but only in the largest
scales\footnote{The same problem arises in the kinetic
theory of gases. We are not interested to know the position and the
velocity of all the molecules of the gas but only some averaged
quantities like the temperature or the pressure. The collisions 
between molecules lead to some mixing at small
scales (molecular chaos), and
this is precisely why the velocities achieve the universal
Maxwell-Boltzmann distribution (most probable state). Similarly, the
statistical theory of 2D turbulence attempts to interpret the coherent
structures (vortices and jets) in terms of appropriate Boltzmann
distributions corresponding to the most probable or most mixed state
of the 2D Euler equation.}. Indeed, the observations are always
realized with a finite resolution. It is therefore desirable to have
an equation for the coarse-grained vorticity $\overline{\omega}({\bf
r},t)$ that parametrizes at best the small scales and that describes
the evolution of the large scales only. Usual parametrizations
introduce a turbulent viscosity in the Euler equations in order to
smooth out the small-scale filaments and prevent numerical
instabilities. However, this artificial viscosity breaks the
conservation of energy. It is therefore important to consider more
general parametrizations of 2D turbulence that smooth out the small
scales while conserving the energy.

In relation to the statistical theory of the 2D Euler equation, some
relaxation equations have been proposed by Robert and Sommeria (1992)
based on a phenomenological Maximum Entropy
Production Principle (MEPP). These equations are constructed so as to
increase the mixing entropy  ($H$-theorem) while conserving all the invariants of the
inviscid dynamics (the energy and the Casimirs) and to relax towards the
maximum entropy state. These equations provide a thermodynamical
parametrization of 2D turbulence. In this approach, the energy is
conserved globally thanks to a uniform Lagrange multiplier $\beta(t)$
which has the interpretation of a global inverse temperature.

In this paper, we propose to apply the MEPP by enforcing a {\it local}
conservation of energy. This leads to an equation for the vorticity
distribution that conserves all the Casimirs, the energy, and that
increases monotonically the mixing entropy. However, this equation
does not generally converge towards the statistical equilibrium state
of the 2D Euler equation. The equation for the coarse-grained
vorticity dissipates monotonically all the generalized enstrophies and
the vorticity current vanishes for {\it any} steady state of the 2D
Euler equation\footnote{We do not know whether this
property is a drawback of the derived equation or if it can account
for (observed) situations where the QSS is different from the
statistical equilibrium state.}. The steady state
that is selected is non-universal. It is determined by the
dynamics in a non-trivial manner
(we have to solve a dynamical equation).  Interestingly, the equation for the
coarse-grained vorticity obtained from the MEPP turns out to coincide
with a particular case of equations obtained by Sadourny and
Basdevant (1981) with the
anticipated vorticity method. Our approach therefore reveals an
interesting connection between the anticipated vorticity method and
the MEPP when the energy conservation is treated locally.

\section{Statistical theory of the 2D Euler equation}
\label{sec_mrs}

Two-dimensional incompressible and inviscid flows are described by the
2D Euler-Poisson system
\begin{equation}
{\partial \omega\over\partial t}+{\bf u}\cdot \nabla \omega=0,\qquad \omega=-\Delta\psi,
\label{mrs1}
\end{equation}
where $\omega$ is the vorticity and $\psi$ the streamfunction. They
are related to the velocity field ${\bf u}$ by $\nabla\times {\bf
u}=\omega{\bf z}$ and ${\bf u}=-{\bf z}\times\nabla\psi$, where ${\bf
z}$ is a unit vector normal to the flow.  Starting from a generically
unsteady or unstable initial condition, the 2D Euler-Poisson system is
known to undergo a complicated mixing process. The vorticity
$\omega({\bf r},t)$ develops a filamentation at smaller and smaller
scales and never reaches a steady state. However, if we locally
average over these filaments, the resulting coarse-grained vorticity
$\overline{\omega}({\bf r},t)$ is expected to reach a quasistationary
state (QSS). This is known as weak convergence in mathematics. This
QSS, which is a particular steady state of the 2D Euler equation,
usually has the form of a large-scale vortex or a jet. A nice
illustration of this mixing process is given by Sommeria {\it et al} (1991)
in connection to the nonlinear development of the Kelvin-Helmholtz
instability in a shear layer.

In order to predict the structure of these QSSs as a function of the
initial condition, Miller (1990) and Robert and Sommeria
(1991) have proposed a statistical theory of the 2D Euler
equation. The key idea is to replace the deterministic description of
the flow $\omega({\bf r},t)$ by a probabilistic description where
$\rho({\bf r},\sigma,t)$ gives the probability density of finding the
vorticity level $\omega=\sigma$ in ${\bf r}$ at time $t$. It satisfies
the normalization condition $\int\rho \, d\sigma=1$. The observed
(coarse-grained) vorticity field is then expressed as
$\overline{\omega}({\bf r},t)=\int \rho\sigma d\sigma$.

To apply the statistical theory, we first have to specify the constraints. The 2D Euler equation conserves the energy
\begin{equation}
\label{mrs3}
E={1\over 2}\int
\overline{\omega}\psi d{\bf r}={1\over 2}\int
\rho\sigma\psi \, d{\bf r}d\sigma
\end{equation}
and the fine-grained vorticity distribution
\begin{equation}
\gamma(\sigma)=\int \rho({\bf
r},\sigma)\, d{\bf r},
\label{mrs4}
\end{equation}
where $\gamma(\sigma)$ is the total area occupied by the
vorticity
level $\sigma$. This is equivalent to the conservation of the
Casimirs $I_{h}=\int
\overline{h(\omega)}d{\bf r}$ where $h$ is an arbitrary function of the vorticity.

The basic object of the statistical theory is the  mixing entropy
\begin{equation}
\label{mrs6}
S[\rho]=-\int
\rho({\bf r},\sigma)\ln \rho({\bf r},\sigma) \, d{\bf r}d\sigma
\end{equation}
which counts the number of microstates corresponding to a given
macrostate (Robert and Sommeria 1991, Chavanis 2002).  The statistical equilibrium state of
the 2D Euler equation, which corresponds to the most probable state
(i.e. the macrostate that is the most represented at the microscopic
level), is obtained by maximizing the mixing entropy (\ref{mrs6})
while respecting the normalization condition $\int\rho \, d\sigma=1$
and conserving all the inviscid invariants of the 2D Euler
equation. If the evolution is ergodic, or at least if mixing is
efficient enough, the system will evolve towards the statistical
equilibrium state of the 2D Euler equation. It is determined by the
maximization problem
\begin{eqnarray}
\label{mrs7}
\max_{\rho}\quad \lbrace S[\rho]\quad | \quad E[\overline{\omega}]=E, \quad
\int\rho({\bf r},\sigma)\, d{\bf r}=\gamma(\sigma), \quad \int \rho \, d\sigma=1\rbrace.
\end{eqnarray}
The critical points of the mixing entropy at fixed $E$, $\gamma(\sigma)$ and normalization are obtained from the variational principle
\begin{equation}
\label{mrs8}
\delta
S-\beta\delta E-\int\alpha(\sigma)\delta\gamma(\sigma)\, d\sigma-\int
\zeta({\bf r})\delta\left (\int \rho \, d\sigma \right ) d{\bf r}=0,
\end{equation}
where $\beta$ (inverse temperature), $\gamma(\sigma)$ (chemical
potential) and $\zeta({\bf r})$ are appropriate Lagrange multipliers. This leads to the equilibrium
distribution
\begin{equation}
\label{mrs9}
\rho({\bf r},\sigma)={1\over Z(\psi({\bf r}))}g(\sigma)  e^{-\beta\sigma\psi({\bf r})},
\end{equation}
where $Z(\psi)=\int g(\sigma) e^{-\beta\sigma\psi}d\sigma$ is
the normalization factor. The coarse-grained vorticity is then given by
\begin{eqnarray}
\label{mrs10}
\overline{\omega}=\frac{\int g(\sigma)\sigma e^{-\beta\sigma\psi} \, d\sigma}{\int g(\sigma) e^{-\beta\sigma\psi}\, d\sigma}=-\frac{1}{\beta}\frac{d\ln Z}{d\psi}=f_{\beta,g}(\psi).
\end{eqnarray}
Differentiating equation (\ref{mrs10}) with respect to $\psi$, it is easy to show that the local centered variance
of the vorticity distribution
\begin{equation}
\label{mrs11}
\omega_2\equiv \overline{\omega^2}-\overline{\omega}^2=\int \rho(\sigma-\overline{\omega})^2\, d{\bf r}
\end{equation}
is given by
\begin{equation}
\label{mrs12}
\omega_{2}=-\frac{1}{\beta}\overline{\omega}'(\psi)=\frac{1}{\beta^2}\frac{d^2\ln Z}{d\psi^2}.
\end{equation}
This relation is reminiscent of the fluctuation-dissipation theorem in statistical mechanics.
Since $\overline{\omega}=\overline{\omega}(\psi)$, the statistical
theory predicts that the coarse-grained vorticity
$\overline{\omega}({\bf r})$ is a {stationary solution} of the 2D
Euler equation. On the other hand, since
$\overline{\omega}'(\psi)=-\beta\omega_{2}(\psi)$ with $\omega_2\ge
0$, the $\overline{\omega}-\psi$ relationship is a {monotonic}
function that is increasing at {negative temperatures} $\beta<0$ and
decreasing at positive temperatures $\beta>0$. Therefore, the
statistical theory {predicts} that the QSS (assumed to be the most
mixed state) is characterized by a monotonic
$\overline{\omega}(\psi)$ relationship. This $\overline{\omega}-\psi$
relationship can take different shapes depending on the initial
condition. Substituting equation (\ref{mrs10}) in the Poisson equation
(\ref{mrs1}-b), the statistical equilibrium state is obtained by solving
the differential equation
\begin{equation}
\label{mrs13}
-\Delta\psi=f_{\beta,g}(\psi)
\end{equation}
with adequate boundary conditions, and relating the Lagrange
multipliers $\beta$ and $g(\sigma)$ to the constraints $E$ and
$\gamma(\sigma)$. We also have to make sure that the distribution
(\ref{mrs9}) is a (local) maximum of entropy, not a minimum or a
saddle point.

\section{Inviscid selective decay and generalized enstrophies}

The moments of the fine-grained vorticity $\Gamma_{n}^{f.g.}=\int
\overline{\omega^{n}}d{\bf r}=\int \rho\sigma^{n}\, d{\bf r}d\sigma$ are conserved since they are particular Casimirs.
The first moment is the circulation $\Gamma=\int
\overline{\omega}d{\bf r}=\int\rho\sigma\, d{\bf r}d\sigma$ and
the second moment is the fine-grained enstrophy $\Gamma_2^{f.g.}=\int
\overline{\omega^2}\, d{\bf r}=\int\rho\sigma^2\, d{\bf r}d\sigma$.
For $n>1$ the moments of the coarse-grained vorticity
$\Gamma_{n>1}^{c.g}=\int
\overline{\omega}^{n}d{\bf r}$ are {\it not} conserved since
$\overline{\omega^{n}}\neq \overline{\omega}^{n}$ (part of the
coarse-grained moments goes into fine-grained fluctuations). For
example, using the Schwarz inequality, we find that
$\Gamma_{2}^{c.g.}=\int
\overline{\omega}^{2}\, d{\bf r}\le \Gamma_{2}^{f.g.}=\int
\overline{\omega^{2}}\, d{\bf r}$. Therefore, the enstrophy
calculated with the coarse-grained vorticity is always smaller than
its initial value while the enstrophy calculated with the fine-grained
vorticity is conserved. Therefore, the notion of coarse-graining
explains how we can have a decrease of enstrophy in a purely inviscid
theory. By contrast, the circulation and the energy calculated with
the coarse-grained vorticity are approximately conserved.  For that reason, the
energy and the circulation are called robust invariants while the
moments of the vorticity of order $n>1$ are called
fragile invariants.  These results may be viewed as a form of inviscid
selective decay due to coarse-graining (not to viscosity).

We introduce a family of functionals of the coarse-grained vorticity of the form
\begin{equation}
\label{g2}
S[\overline{\omega}]=-\int C(\overline{\omega})\, d{\bf r},
\end{equation}
where $C$ is any convex function (i.e. $C''>0$). It can be shown that
these functionals increase with time (due to mixing) in the sense that
$S(t)\ge S(0)$ for any $t>0$. For that reason they are sometimes
called ``generalized $H$-functions'' (Tremaine {\it et al} 1986, Appendix A
of Chavanis 2006) or ``generalized entropies'' (Chavanis
2003). However, (i) a monotonic increase of $S(t)$
(i.e. an $H$-theorem) and (ii) the relaxation of the
system towards a maximum of one of these functionals $S$ are
{\it not} implied by this theorem (although these properties may be
expected in generic situations). We note that the neg-enstrophy (the
opposite of the enstrophy $\Gamma_2^{c.g.}=\int \overline{\omega}^2\,
d{\bf r}$) is a particular case of such functionals. For that reason,
the functionals $-S$ are sometimes called ``generalized enstrophies''.

\section{The Maximum Entropy Production Principle with a global conservation of energy}
\label{sec_mepp}

Let us decompose the vorticity $\omega$ and the
velocity ${\bf u}$ into a mean and a fluctuating part, namely
$\omega=\overline{\omega }+\tilde{\omega }$,
${\bf{u}}=\overline{\bf{u}}+\tilde{\bf{u}}$. Taking the local average
of the Euler equation (\ref{mrs1}-a), we get
\begin{equation}
\label{mepp1}
{\partial \overline{\omega} \over \partial t}+\nabla\cdot (\overline{\omega}\, \overline{\bf  u})=-\nabla \cdot {\bf  J}_{\omega},
\end{equation}
where the vorticity current $\bf{J}_{\omega }=\overline{\tilde{\omega
}\tilde{\mathbf{u}}}$ represents the correlations of the fine-grained
fluctuations. Equation (\ref{mepp1}) can be viewed as a local conservation
law for the circulation $\Gamma=\int
\overline{\omega}\, d{\bf{r}}$. In order to conserve all the Casimirs, we need to
consider not only the locally averaged vorticity field
$\overline{\omega}({\bf{r}},t)$ but the whole probability distribution $\rho
({\bf{r}},{\sigma },t)$ now evolving with time $t$.  The conservation
of the global vorticity distribution $\gamma (\sigma )=\int \rho\,
d{\bf{r}}$ can be written in the local form as
\begin{equation}
\label{mepp2}
{\partial \rho\over \partial t}+\nabla\cdot (\rho\overline{\bf  u})=-\nabla \cdot {\bf  J},
\end{equation}
where ${\bf J}({\bf r},\sigma,t)$ is the (unknown) current
associated with the vorticity level $\sigma$. Integrating
equation (\ref{mepp2}) over all the vorticity levels $\sigma$, we find the
constraint $\int {\bf{J}}({\bf{r}},\sigma ,t)\, d\sigma ={\bf 0}$. This
accounts for the conservation of the normalization condition $\int\rho\, d\sigma=1$. Multiplying
equation (\ref{mepp2}) by $\sigma$ and integrating over all the vorticity
levels, we get $\int {\bf{J}}({\bf{r}},\sigma ,t)\sigma \, d\sigma
={\bf{J}}_{\omega }$.

We can express the time variation of energy in terms of ${\bf J}_{\omega}$, using equations (\ref{mrs3})
and (\ref{mepp1}). This leads to the constraint
\begin{eqnarray}
\label{mepp3}
\dot E=\int {\bf J}_{\omega}\cdot \nabla\psi \, d{\bf r}=0.
\end{eqnarray}
Using equations (\ref{mrs6}) and (\ref{mepp2}), we similarly express the
rate of entropy production as
\begin{equation}
\label{mepp4}
\dot{S}=-\int {\bf{J}}\cdot  \frac{\nabla\rho}{\rho} \, d{\bf{r}}d\sigma .
\end{equation}

The Maximum Entropy Production Principle (MEPP) consists in choosing
the current ${\bf{J}}$ which maximizes the rate of entropy production
$\dot{S}$ respecting the constraints
$\dot{E}=0$, $\int {\bf
J}\, d\sigma={\bf 0}$, and $\int {J^{2}\over 2\rho }\, d\sigma \leq
C({\mathbf{r}},t)$.  The last constraint expresses a bound (unknown)
on the value of the diffusion current. Convexity arguments justify
that this bound is always reached so that the inequality can be
replaced by an equality. The corresponding condition on first
variations can be written at each time $t$:
\begin{eqnarray}
\label{mepp5}
\delta \dot S-\beta(t)\delta \dot E-\int \mathbf{\zeta} ({\bf r},t)\delta \biggl (\int {\mathbf{J}}d\sigma \biggr )\, d{\mathbf{r}}-\int \frac{1}{D({\mathbf{r}},t)}\delta \biggl (\int {{\bf J}^{2}\over 2\rho} d\sigma\biggr ) \, d{\mathbf{r}}=0\quad
\end{eqnarray}
where $\beta(t)$, $\zeta({\bf r},t)$ and $D({\bf r},t)$ are Lagrange multipliers, and leads to a current of the form
\begin{equation}
\label{mepp6}
{\bf J}=-D({\bf r},t)\left\lbrack \nabla\rho+\beta(t)\rho(\sigma-\overline{\omega})\nabla\psi\right\rbrack.
\end{equation}
The Lagrange multiplier ${\zeta}({\bf r},t)$ has been eliminated, using the
condition $\int {\bf
J}\, d\sigma={\bf 0}$ of local normalization conservation.
The vorticity current is
\begin{equation}
\label{mepp7}
{\bf J}_{\omega}=-D({\bf r},t)\left\lbrack \nabla\overline{\omega}+\beta(t)\omega_2\nabla\psi\right\rbrack.
\end{equation}
The thermodynamical parametrization proposed by Robert and Sommeria (1992) can therefore be written as (for simplicity we do not write the bar on ${\bf u}$):
\begin{equation}
\label{mepp8}
{\partial \rho\over \partial t}+{\bf u}\cdot\nabla\rho=\nabla \cdot \biggl\lbrace D({\bf r},t) \biggl\lbrack\nabla\rho+\beta(t)\rho(\sigma-\overline{\omega})\nabla\psi\biggr\rbrack\biggr\rbrace.
\end{equation}
The equation for the coarse-grained vorticity is
\begin{equation}
\label{mepp9}
{\partial \overline{\omega} \over \partial t}+{\bf u}\cdot\nabla\overline{\omega}=\nabla \cdot \biggl\lbrace D({\bf r},t) \biggl\lbrack\nabla\overline{\omega}+\beta(t)\omega_2\nabla\psi\biggr\rbrack\biggr\rbrace.
\end{equation}
We note that this equation is not closed since it depends on the local centered enstrophy $\omega_2$. We therefore have to solve equation (\ref{mepp8}) for all the levels or write an infinite hierarchy of equations for the moments $\overline{\omega^k}$ (Robert and Rosier 1997). The time evolution of the Lagrange multiplier $\beta(t)$ is determined by introducing equation (\ref{mepp7}) in the constraint (\ref{mepp3}). This yields
\begin{equation}
\label{mepp10}
\beta (t)=-{\int D\nabla \overline{\omega}\cdot \nabla\psi \, d{\bf{r}}\over \int D\omega_{2}(\nabla\psi)^{2}\, d{\bf{r}}}.
\end{equation}

The mixing entropy (\ref{mrs6}) satisfies an $H$-theorem provided that $D\ge 0$. Indeed, using the expression (\ref{mepp6}) of the current, the entropy production (\ref{mepp4}) can be rewritten as
\begin{eqnarray}
\label{mepp11}
\dot{S}=\int {{\bf J}^{2}\over D\rho }\, d{\bf{r}}d\sigma+\int\frac{\bf J}{\rho}\cdot \left\lbrace \beta(t)\rho(\sigma-\overline{\omega})\nabla\psi\right\rbrace\, d{\bf{r}}d\sigma.
\end{eqnarray}
Integrating over the vorticity levels in the second term, and using the conservation of energy (\ref{mepp3}), we get
\begin{equation}
\label{mepp12}
\dot{S}=\int {{\bf J}^{2}\over D\rho }\, d{\bf{r}}d\sigma+\beta(t)\int {\bf J}_{\omega}\cdot \nabla\psi\, d{\bf{r}}=\int {{\bf J}^{2}\over D\rho }\, d{\bf{r}}d\sigma\ge 0.
\end{equation}
A stationary solution of equation (\ref{mepp8}) satisfies $\dot{S}=0$ implying ${\bf{J}}={\bf 0}$. Using equation (\ref{mepp6}),
we obtain $\nabla\ln \rho+\beta (\sigma -\overline{\omega})\nabla
\psi={\bf 0}$.  For any reference vorticity level $\sigma_{0}$,
it writes $\nabla\ln \rho _{0}+\beta (\sigma _{0}-\overline{\omega})\nabla \psi={\bf 0}$. Subtracting the foregoing equations, we obtain $\nabla
\ln ({\rho /\rho _{0}})+\beta (\sigma -\sigma _{0})\nabla \psi ={\bf
0}$, which is immediately integrated into equation (\ref{mrs9}) where
$Z^{-1}({\bf{r}})\equiv \rho _{0}({\bf{r}})e^{\beta
\sigma _{0}\psi ({\bf{r}})}$ and $g(\sigma )\equiv e^{A(\sigma
)}$, $A(\sigma )$ being a constant of integration. Therefore, the
mixing entropy (\ref{mrs6}) increases monotonically until the
distribution (\ref{mrs9}) is reached, with $\beta=\lim _{t\rightarrow
\infty }\beta (t)$. It can be shown that a stationary solution of the
relaxation equation (\ref{mepp8}) is linearly stable if, and only if,
it is an entropy {\it maximum} at fixed energy and
Casimirs. Therefore, this numerical algorithm selects the maxima (and
not the minima or the saddle points) among all the critical points of
entropy. When several entropy maxima subsist for the same values of
the constraints, the choice of equilibrium depends on a complicated
notion of ``basin of attraction'' and not simply whether the solution
is a local or a global entropy maximum.

In summary, equation (\ref{mepp8}) conserves the energy (\ref{mrs3}), the Casimirs (\ref{mrs4}), and
increases monotonically the mixing entropy (\ref{mrs6}) ($H$-theorem).  For
$t\rightarrow +\infty$, the solution converges towards the maximum entropy state
(\ref{mrs9}). The generalized enstrophies (\ref{g2}) are not conserved but it does not seem possible to prove whether they decay monotonically or not  (actually, there is no fundamental reason why they should decay monotonically in an inviscid theory).

The vorticity current in the relaxation equation (\ref{mepp9}) is the
sum of two terms. A term ${\bf J}_{diff}=-D\nabla\overline{\omega}$
leading to a pure diffusion with a turbulent viscosity $D$ and an
additional term ${\bf J}_{drift}=-\beta D\omega_2\nabla\psi$
interpreted as a {\it drift}. The drift coefficient (mobility) is
given by a sort of Einstein relation\footnote{There exist numerous
analogies between the kinetic theories of 2D vortices and stellar
systems. In these analogies, the drift of the vortices is the
counterpart of the dynamical friction experienced by a star (Chavanis
2002).}. The relaxation equation (\ref{mepp9}) may be interpreted as a
nonlinear Fokker-Planck equation. It shares some analogies with the
Fokker-Planck equation obtained in the kinetic theory of point
vortices in the thermal bath approximation (Chavanis 2001, 2002)
although the physics of the problem is fundamentally different. On the
other hand, a kinetic equation for the coarse-grained vorticity has
been derived from a quasilinear theory of the 2D Euler equation
(Chavanis 2000, 2002), and some connections with the
relaxation equations issued from the MEPP have been mentioned.

Usual parameterizations of 2D turbulence  include a single turbulent viscosity. In a sense, they correspond to the infinite temperature limit ($\beta=0$) of the relaxation equation (\ref{mepp9}) where the drift vanishes. However, these equations without the drift term do not conserve the energy. The drift term is therefore necessary to restore this property.

\section{The Maximum Entropy Production Principle with a local conservation of energy}
\label{sec_l}

In the previous section, the energy is conserved globally thanks to a uniform Lagrange multiplier $\beta(t)$ interpreted as a global inverse temperature. In this section, we propose to apply the MEPP by imposing a local conservation of energy ${\bf J}_{\omega}\cdot \nabla\psi =0$. In that case, the variational problem (\ref{mepp5}) is replaced by
\begin{eqnarray}
\label{l1}
\delta \dot S-\int \beta({\bf r},t)\delta ({\bf J}_{\omega}\cdot \nabla\psi)\, d{\bf r}
-\int {\bf \zeta} ({\bf r},t) \delta \biggl (\int {\mathbf{J}}d\sigma \biggr )\, d{\mathbf{r}}\nonumber\\
-\int \frac{1}{D({\mathbf{r}},t)}\delta \biggl (\int {{\bf J}^{2}\over 2\rho} d\sigma\biggr ) \, d{\mathbf{r}}=0.
\end{eqnarray}
It leads to an optimal current of the form
\begin{equation}
\label{l2}
{\bf J}=-D({\bf r},t)\left\lbrack \nabla\rho+\beta({\bf r},t)\rho(\sigma-\overline{\omega})\nabla\psi\right\rbrack.
\end{equation}
The vorticity current is
\begin{equation}
\label{l3}
{\bf J}_{\omega}=-D({\bf r},t)\left\lbrack \nabla\overline{\omega}+\beta
({\bf r},t)\omega_2\nabla\psi\right\rbrack.
\end{equation}
The evolution of the Lagrange multiplier $\beta({\bf r},t)$ is determined
by introducing Eq. (\ref{l3}) in the local energy constraint ${\bf J}_{\omega}\cdot \nabla\psi =0$. This yields $\nabla\psi\cdot\nabla\overline{\omega}+\beta\omega_2(\nabla\psi)^2=0$ implying
\begin{equation}
\label{l4}
\beta({\bf r},t)=-\frac{\nabla\psi\cdot\nabla\overline{\omega}}{\omega_2(\nabla\psi)^2}.
\end{equation}
Substituting these expressions in equations (\ref{mepp1}) and (\ref{mepp2}), we obtain the parametrization
\begin{equation}
\label{l5}
{\partial \rho\over \partial t}+{\bf u}\cdot\nabla\rho=\nabla \cdot \left\lbrace D({\bf r},t) \left\lbrack\nabla\rho-\rho(\sigma-\overline{\omega})
\frac{\nabla\psi\cdot\nabla\overline{\omega}}{\omega_2(\nabla\psi)^2}\nabla\psi\right\rbrack\right\rbrace
\end{equation}
and
\begin{equation}
\label{l6}
{\partial \overline{\omega} \over \partial t}+{\bf u}\cdot\nabla\overline{\omega}=\nabla \cdot \left\lbrace D({\bf r},t) \left \lbrack\nabla\overline{\omega}-\frac{\nabla\psi\cdot\nabla\overline{\omega}}{(\nabla\psi)^2}\nabla\psi\right \rbrack\right\rbrace.
\end{equation}
Since the diffusion coefficient is unspecified, we can take $D\propto (\nabla\psi)^2$ to avoid dividing by zero when $\nabla\psi={\bf 0}$.

These equations conserve the normalization, the total surface of each vorticity level, and the energy. They also increase the mixing entropy (\ref{mrs6}) monotonically provided that $D\ge 0$. Indeed, using the expression (\ref{l2}) of the current, the entropy production (\ref{mepp4}) can be rewritten as
\begin{equation}
\label{l7}
\dot{S}=\int {{\bf J}^{2}\over D\rho }\, d{\bf{r}}d\sigma+\int\frac{\bf J}{\rho}\cdot \left\lbrack \beta({\bf r},t)\rho(\sigma-\overline{\omega})\nabla\psi\right\rbrack\, d{\bf{r}}d\sigma.
\end{equation}
Integrating over the vorticity levels, and using the local conservation of energy, we get
\begin{equation}
\label{l8}
\dot{S}=\int {{\bf J}^{2}\over D\rho }\, d{\bf{r}}d\sigma+\int \beta({\bf r},t) {\bf J}_{\omega}\cdot \nabla\psi\, d{\bf{r}}=\int {{\bf J}^{2}\over D\rho }\, d{\bf{r}}d\sigma\ge 0.
\end{equation}
It is also possible to prove that equation (\ref{l6}) for the coarse-grained vorticity dissipates all the generalized enstrophies (\ref{g2}) monotonically provided that $D\ge 0$. The rate of dissipation of the generalized enstrophies is the opposite of
\begin{equation}
\label{l9}
\dot S=-\int C''(\overline{\omega}) {\bf J}_{{\omega}}\cdot \nabla\overline{\omega} \, d{\bf r}.
\end{equation}
Using the expression (\ref{l3}) of the vorticity current and the local conservation of energy, we get
\begin{equation}
\label{l10}
\dot{S}=\int C''(\omega)\frac{{\bf J}_{\omega}^{2}}{D}\, d{\bf{r}}+\int C''(\overline{\omega})\beta\omega_2 {\bf J}_{\omega}\cdot\nabla\psi\, d{\bf{r}}=\int C''(\omega){{\bf J}_{\omega}^{2}\over D}\, d{\bf{r}}\ge 0.
\end{equation}

In summary, equation (\ref{l5}) conserves the energy (\ref{mrs3}), the
Casimirs (\ref{mrs4}), and increases monotonically the mixing entropy
(\ref{mrs6}) ($H$-theorem). We emphasize, however, that this equation
does {\it not} relax towards the maximum entropy state in
general. Indeed, equation (\ref{l6}) for the coarse-grained vorticity
dissipates all the generalized enstrophies (\ref{g2})
monotonically. Furthermore, the vorticity current vanishes for {\it
any} steady state of the 2D Euler equation such that
$\overline{\omega}=f(\psi)$. Therefore, the system is expected to
reach a steady state of the 2D Euler equation but its precise form
cannot be determined {\it a priori}. It is non-universal. It
depends
on the dynamics and we have to solve Eq. (\ref{l6}).

We note the remarkable fact that the equation for the coarse-grained vorticity (\ref{l6}) is {\it closed}. This is not the case in the parametrization (\ref{mepp9}) where it depends on the local enstrophy $\omega_2({\bf r},t)$. This is a great practical advantage of the present parametrization since we do not have to solve the equation for the vorticity distribution (\ref{l5}), or consider an infinite hierarchy of moments equations, to obtain the evolution of the coarse-grained  vorticity (which is the quantity of main interest)\footnote{The fact that the results depend on the detailed vorticity distribution $\rho({\bf r},\sigma,t)$, as implied by the statistical theory of 2D turbulence, is the main practical difficulty to implement the parametrization (\ref{mepp8}). In realistic applications, it is difficult to determine what are the levels to consider.}. This ``miracle'' only occurs for the   coarse-grained  vorticity. The evolution of the higher order moments $\overline{\omega^k}({\bf r},t)$ are given by an infinite hierarchy of equations obtained from equation (\ref{l5}).

\section{The anticipated vorticity method}
\label{sec_av}

In the presence of a very small viscosity, the energy and the
circulation are almost conserved while the enstrophy decays
monotonically (actually, all the generalized enstrophies decay
monotonically, see Appendix A of Chavanis 2006). This property of
selective decay has led to the minimum enstrophy
principle\footnote{This ``principle'' is only
phenomenological. Indeed, the conservation of energy and circulation,
and the monotonic decay of enstrophy, do not guarantee that the system
will necessarily reach a minimum enstrophy state at fixed energy and
circulation.}. It also suggests to develop a parametrization that
conserves the energy and the circulation while dissipating
monotonically the enstrophy (or the generalized enstrophies). These
considerations have lead to the parametrization of Sadourny and
Basdevant (1981) based on the anticipated vorticity method.

Following Sadourny and Basdevant (1981), we determine the vorticity current ${\bf
J}_{\omega}$ in order to conserve locally the energy and decrease monotonically all the generalized enstrophies (Sadourny and Basdevant only consider the dissipation of enstrophy but we show below that, actually, all the generalized enstrophies decay). We assume that the energy is
conserved locally so that ${\bf J}_{\omega}\cdot\nabla\psi=0$. This implies
that the vorticity current must be parallel to the velocity, i.e.
${\bf J}_{\omega}=-\lambda({\bf r},t){\bf u}$ where $\lambda({\bf r},t)$ is an arbitrary function. Substituting this relation in equation (\ref{l9}) we get $\dot S=\int
C''(\overline{\omega})\lambda({\bf r},t){\bf u} \cdot
\nabla\overline{\omega} \, d{\bf r}$.  If we take $\lambda({\bf
r},t)=K({\bf r},t){\bf u} \cdot \nabla\overline{\omega}$ with $K\ge
0$, we obtain $\dot S\ge 0$. Finally, it is relevant to write  $K=D/u^2$ where
$D({\bf r},t)\ge 0$ has the dimension of a diffusion
coefficient. Therefore ${\bf J}_{\omega}=-D({\bf
u}\cdot\nabla\overline{\omega}){\bf u}/u^2$. Substituting this expression in equation (\ref{mepp1}), we obtain
\begin{equation}
\label{av1}
{\partial \overline{\omega} \over \partial t}+{\bf  u}\cdot \nabla\overline{\omega}=\nabla\cdot \left\lbrack D \frac{({\bf u}\cdot \nabla\overline{\omega}){\bf u}}{u^2}\right\rbrack.
\end{equation}
This equation can also be written as
\begin{equation}
\label{av2}
{\partial \overline{\omega} \over \partial t}+{\bf  u}\cdot \nabla\overline{\omega}=\nabla\cdot \left (D \frac{{\bf u}\otimes {\bf u}}{u^2}\nabla\overline{\omega}\right ).
\end{equation}
Since $D$ is unspecified, we can take $D=Ku^2$
in order to avoid dividing by zero when $u=0$. Equation (\ref{av1}) is a particular case of the parametrization proposed by Sadourny and Basdevant (1981). This equation conserves locally the energy and dissipates monotonically all
the generalized enstrophies:
\begin{equation}
\label{av3}
\dot S=\int C''(\overline{\omega})\frac{D}{u^2}({\bf u}\cdot \nabla\overline{\omega})^2 \, d{\bf r}\ge 0.
\end{equation}
The diffusion current vanishes when a steady state of the 2D Euler
equation is reached.  Indeed, ${\bf J}_{\omega}=0$ when ${\bf u}\cdot
\nabla\overline{\omega}=0$ which is equivalent to
$\overline{\omega}=f(\psi)$. However, it does not seem possible to
predict that steady state {\it a priori}. Its selection depends
on the dynamics in a
non-trivial manner and we have to solve equation (\ref{av1}).

Finally, we show that equation (\ref{av1}) is equivalent to equation (\ref{l6}) derived from the MEPP. Combining the identity of vector analysis
\begin{equation}
\label{av4}
{\bf u}\times (\nabla\overline{\omega}\times {\bf u})=u^2\nabla\overline{\omega}-({\bf u}\cdot \nabla\overline{\omega}){\bf u}
\end{equation}
with the relation
\begin{equation}
\label{av5}
\nabla\overline{\omega}\times {\bf u}=\nabla\overline{\omega}\times (-{\bf z}\times\nabla\psi)=-(\nabla\overline{\omega}\cdot\nabla\psi){\bf z}
\end{equation}
leading to
\begin{equation}
\label{av6}
{\bf u}\times (\nabla\overline{\omega}\times {\bf u})=(\nabla\overline{\omega}\cdot\nabla\psi)\nabla\psi,
\end{equation}
we find that
\begin{equation}
\label{av7}
\frac{({\bf u}\cdot \nabla\overline{\omega}){\bf u}}{u^2}=\nabla\overline{\omega}-\frac{(\nabla\overline{\omega}\cdot\nabla\psi)\nabla\psi}{(\nabla\psi)^2}.
\end{equation}
Therefore, equation (\ref{av1}) is the same as equation (\ref{l6}).

\section{Selective decay principle}
\label{sec_selective}

As we have seen, equation (\ref{av1}) is equivalent to equation
(\ref{l6}). However, the form (\ref{l6}) of this equation, which does
not seem to have been noticed before, has a very interesting
structure. The right hand side of equation (\ref{l6}) is the sum of two
terms. A diffusion term and a drift term. Usual parameterizations only
include a diffusion term. However, this diffusion term alone
dissipates the energy which is a bad feature of these
parametrizations. We show below that the drift term acts precisely in
a way to restore the conservation of energy. Indeed, we can write
$\dot E=\dot E_{diff}+\dot E_{drift}$ with
\begin{equation}
\label{sd1}
\dot E_{diff}=\int {\bf J}_{diff}\cdot\nabla\psi\, d{\bf r}=-\int D \nabla\overline{\omega}\cdot\nabla\psi\, d{\bf r}=-D\int\overline{\omega}^2\, d{\bf r}\le 0,
\end{equation}
\begin{equation}
\label{sd2}
\dot E_{drift}=\int {\bf J}_{drift}\cdot\nabla\psi\, d{\bf r}=\int D \frac{\nabla\psi\cdot\nabla\overline{\omega}}{(\nabla\psi)^2}\nabla\psi\cdot\nabla\psi\, d{\bf r}=D\int\overline{\omega}^2\, d{\bf r}\ge 0.
\end{equation}
In order to obtain the last integral, we have assumed that $D$ is constant and used an integration by parts (but this last step is not necessary for the proof). We see that $\dot E_{drift}=-\dot E_{diff}\ge 0$. The diffusion term dissipates the energy while the drift term increases it. As a whole, the energy is conserved: $\dot E=0$.

A pure diffusion term dissipates the generalized enstrophies
monotonically (see Appendix A of Chavanis (2006)). We show below that
this property persists in the presence of the drift term. Indeed, we
can write $\dot S=\dot S_{diff}+\dot S_{drift}$ with
\begin{equation}
\label{sd3}
\dot S_{diff}=-\int C''(\overline{\omega}) {\bf J}_{diff}\cdot\nabla\overline{\omega}\, d{\bf r}=\int  D C''(\overline{\omega})(\nabla\overline{\omega})^2\, d{\bf r}\ge 0,
\end{equation}
\begin{equation}
\label{sd4}
\dot S_{drift}=-\int C''(\overline{\omega}) {\bf J}_{drift}\cdot\nabla\overline{\omega}\, d{\bf r}=-\int D C''(\overline{\omega}) \frac{(\nabla\psi\cdot\nabla\overline{\omega})^2}{(\nabla\psi)^2}\, d{\bf r}\le 0.
\end{equation}
The diffusion term dissipates the generalized enstrophies while the drift term increases them. As a whole, the generalized enstrophies decay monotonically:  $\dot S\ge 0$.

These properties are strikingly consistent with the phenomenology of
2D turbulence. With only the diffusion term, we have a direct cascade
of enstrophy and a spurious direct cascade of energy (for a
``large'' turbulent viscosity). With the diffusion
term and the drift term, we have a direct cascade of enstrophy and an
inverse cascade of energy. Equation (\ref{l6}) is therefore consistent with
the phenomenological selective decay principle. It dissipates the
generalized enstrophies while conserving the energy.

{\it Remark:} We can obtain similar relations for the parametrization (\ref{mepp8}) associated with the entropy (\ref{mrs6}). We first have $\dot E_{drift}=-\dot E_{diff}=-\int D\nabla\overline{\omega}\cdot\nabla\psi\, d{\bf r}\ge 0$ leading to $\dot E=0$. We also find that $\dot S_{diff}=\int D [(\nabla\rho)^2/\rho]\, d{\bf r}\ge 0$ and $\dot S_{drift}=\beta(t)\dot E_{drift}$. If $\beta(t)\le 0$, then $\dot S_{drift}\le 0$. In any case, $\dot S\ge 0$.

\section{Differences with other equations}

Equation (\ref{l6}) is different from the relaxation equation
\begin{equation}
\label{two1}
\frac{\partial\overline{\omega}}{\partial t}+{\bf u}\cdot \nabla\overline{\omega}=\nabla\cdot \left\lbrack D({\bf r},t)\left (\nabla\overline{\omega}+\frac{\beta(t)}{C''(\overline{\omega})}\nabla\psi\right )\right\rbrack,
\end{equation}
\begin{equation}
\label{two2}
\beta(t)=-\frac{\int D\nabla\overline{\omega}\cdot\nabla\psi\, d{\bf r}}{\int D\frac{(\nabla\psi)^{2}}{C''(\overline{\omega})}\, d{\bf r}}
\end{equation}
derived by Chavanis (2003) from a generalized MEPP in
$\overline{\omega}$-space.
This equation
increases monotonically a {\it particular} generalized entropy $S$,
specified by the convex function $C(\overline{\omega})$, while
conserving energy $E$ and circulation $\Gamma$. It relaxes towards
a (local)
maximum of $S$ at fixed $\Gamma$ and $E$. We note that
equations (\ref{two1}) and (\ref{two2}) may be obtained from
equations (\ref{mepp9}) and (\ref{mepp10}) by making the {\it ansatz}
${\omega}_{2}({\bf r},t)=1/C''[\overline{\omega}({\bf r},t)]$.  We
also note that equation (\ref{l6}) can be obtained from equations  (\ref{two1})
and (\ref{two2}) if the global conservation of energy is replaced by a
local conservation of energy, i.e. if we suppress the integrals in
equation (\ref{two2}). This shows that the decay of the generalized
enstrophies in equation (\ref{l6}) is optimal.

Equation (\ref{l6}) is the ``opposite'' of the equation
\begin{equation}
\label{two3}
\frac{\partial{\omega}}{\partial t}+{\bf u}\cdot \nabla{\omega}=-\alpha\lbrace{\omega},\lbrace{\omega},\psi\rbrace\rbrace
\end{equation}
proposed by Vallis {\it et al} (1989). For $\alpha<0$ this equation dissipates the energy monotonically while
conserving all the Casimirs. It relaxes towards the minimum of energy
under isovortical perturbations.

We refer to Chavanis (2009) for a more detailed discussion of the variational problems of 2D turbulence and the corresponding relaxation equations.

\section{Discussion}
\label{sec_discussion}

Robert and Sommeria (1992) have used a MEPP with a global conservation of energy and
derived the parametrization (\ref{mepp8}). It is possible to take into account the conservation of angular momentum $L=\int \omega r^2\, d{\bf r}$ and linear impulse ${\bf P}=-{\bf z}\times \int\omega {\bf r}\, d{\bf r}$ in their parametrization by introducing
appropriate Lagrange multipliers $\Omega(t)$ and ${\bf U}(t)$ in the variational principle (\ref{mepp5}). In that case, the stream function $\psi({\bf r},t)$ is replaced by the relative stream function $\psi_{eff}({\bf r},t)=\psi({\bf r},t)+\frac{\Omega(t)}{2}r^2-{\bf U}_{\perp}(t)\cdot {\bf r}$ (Chavanis {\it et al} 1996). This parametrization works well to describe the organization of a 2D turbulent flow into a {\it single} coherent structure, for example the vortex resulting from the merging of two vortices (Robert and Sommeria 1992, Robert and Rosier 1997).
However, this parametrization does not respect the galilean invariance of the 2D Euler
equation. In addition, the conservation of energy, angular momentum, and linear
impulse are enforced globally thanks to uniform Lagrange multipliers
$\beta(t)$, $\Omega(t)$ and ${\bf U}(t)$. Not only
this procedure is artificial, but it also poses practical problems to describe
large-scale flows that organize into several {\it distinct} coherent structures.
Indeed, if we view these individual structures as maximum entropy states, there
is no reason why they should all have the same temperature, angular velocity, and linear impulse. In addition, in order to determine $\beta(t)$, $\Omega(t)$ and ${\bf U}(t)$ in the parametrization
(\ref{mepp8}) we have to integrate over the whole domain while the physics of
the problem should be more local, even though the interaction is long-range. For
example, to describe the formation of a large cyclone over a part of the world, it should not
be necessary to perform an integral over the whole sphere (the Earth) as implied
by equation (\ref{mepp10}).  To solve these problems, Chavanis and Sommeria (1997)
have proposed a set of relaxation equations that preserve the galilean
invariance of the 2D Euler equation and that satisfy the conservation of energy,
angular momentum and linear impulse locally thanks to diffusion currents.
These equations relax towards individual
coherent structures that correspond to statistical equilibrium states
(\ref{mrs9}) having {\it different} values of temperature, angular velocity,
and linear impulse. They may be interpreted as ``maximum entropy
bubbles'' (Chavanis and Sommeria 1998). For the moment, this parametrization has
never been used in practice. One difficulty is that we have to consider a large
number of coupled equations for each level $\sigma$ or an infinite hierarchy of equations for the vorticity moments $\overline{\omega^k}({\bf r},t)$ (a difficulty inherent to the
statistical theory of 2D turbulence). In the present paper, we have considered a different approach. We have used the MEPP with a local conservation of energy\footnote{The difference with Chavanis and Sommeria (1998) is that we take here the current of energy ${\bf J}_{\epsilon}$ equal to zero, i.e. we impose ${\bf J}_{\omega}\cdot \nabla\psi=0$ instead of ${\bf J}_{\omega}\cdot \nabla\psi=\nabla\cdot {\bf J}_{\epsilon}$ with ${\bf J}_{\epsilon}\neq {\bf 0}$.} and derived the parametrization (\ref{l5})-(\ref{l6}).  We have shown that this parametrization is equivalent to a special case (\ref{av1}) of the parametrization  of Sadourny and Basdevant (1981) based on the anticipated vorticity method, although the equations appear in a different form [compare equations (\ref{l6}) and (\ref{av1})]. The new form of equation (\ref{l6}) derived in the present paper has a more physical interpretation than the form (\ref{av1}) (see section \ref{sec_selective}). An advantage of this parametrization over the parameterizations of Robert and Sommeria (1992) and Chavanis and Sommeria (1997) is that it yields a closed equation for the coarse-grained vorticity $\overline{\omega}({\bf r},t)$ instead of an infinite hierarchy of equations for the vorticity moments $\overline{\omega^k}({\bf r},t)$. However, it does not respect the conservation of angular momentum and linear impulse, nor the  galilean invariance of the 2D Euler equation. Finally, it does not in general relax towards a maximum entropy state unlike the parameterizations of Robert and Sommeria (1992) and Chavanis and Sommeria (1997). Indeed, in the parametrization (\ref{l6}) or (\ref{av1}) the diffusion current vanishes for any steady state of the 2D Euler equation. This may account for non-ergodicity or be a drawback of this parametrization. Finally, the relaxation equation (\ref{two1})  of Chavanis (2003) is somehow intermediate between these different parameterizations since it relaxes towards a special class of steady states of the 2D Euler equations specified by a generalized entropy $S[\overline{\omega}]$. It is straightforward to generalize this equation in order to conserve energy, angular momentum and linear impulse locally by making the {\it ansatz}
${\omega}_{2}({\bf r},t)=1/C''[\overline{\omega}({\bf r},t)]$ in the parametrization of Chavanis and Sommeria (1997). Of course, it would be interesting to compare the efficiency of these different  parameterizations through direct numerical simulations.

\section{Conclusion}
\label{sec_conclusion}

We have shown a connection between the anticipated vorticity
method of Sadourny and Basdevant (1981) and the Maximum Entropy Production
Principle of Robert and Sommeria (1992) when the conservation of energy is
treated locally (instead of globally). This connection is new and is the main
result of the paper. More than showing the relation between two known equations,
we have derived from the MEPP a new form of equation [see equation (\ref{l6})] that
turns out, after some algebraic manipulations, to coincide with a special case
of the parametrization of Sadourny and Basdevant (1981) [see equation (\ref{av1})].
The new form 
of equation (\ref{l6}), which incorporates a diffusion term and a drift term,
has a nice physical interpretation in terms of a selective decay principle. This
gives a new light to both the MEPP and  the anticipated vorticity method.

\appendix

\section{The equation for the velocity field}
\label{sec_v}

The equation for the coarse-grained velocity field is
\begin{equation}
\label{v1}
\frac{\partial {\bf u}}{\partial t}+({\bf u}\cdot \nabla){\bf u}=-\frac{1}{\rho}\nabla p-{\bf z}\times {\bf J}_{\omega},
\end{equation}
where $p$ is the pressure and ${\bf J}_{\omega}$ the vorticity current. Taking the curl of equation (\ref{v1}) and using the identity $\nabla\times({\bf z}\times {\bf a})=(\nabla\cdot {\bf a}){\bf z}$ and the definition $\nabla\times {\bf u}=\overline{\omega} {\bf z}$ we recover equation (\ref{mepp1}). In section \ref{sec_l}, we have established that
\begin{equation}
\label{v2}
{\bf J}_{\omega}=-D\left (\nabla\overline{\omega}-\frac{\nabla\overline{\omega}\cdot \nabla\psi}{(\nabla\psi)^2}\nabla\psi\right ).
\end{equation}
Using the identity of vector analysis $\Delta{\bf u}=\nabla(\nabla\cdot {\bf u})-\nabla\times(\nabla\times {\bf u})$ which reduces for a 2D incompressible flow to  $\Delta{\bf u}={\bf z}\times\nabla\overline{\omega}$ we obtain
\begin{equation}
\label{v3}
\frac{\partial {\bf u}}{\partial t}+({\bf u}\cdot \nabla){\bf u}=-\frac{1}{\rho}\nabla p+D\left (\Delta{\bf u}+\frac{\nabla\overline{\omega}\cdot \nabla\psi}{(\nabla\psi)^2}{\bf u}\right ).
\end{equation}
Under this form, the drift is equivalent to a force directed along the velocity ${\bf u}$. It points in the same direction as the velocity when $\nabla\overline{\omega}\cdot \nabla\psi>0$ which corresponds to negative temperatures. It may therefore  be interpreted as an anti-friction, or a forcing, that restores the conservation of energy dissipated by the diffusion term. Using the equivalent expression
\begin{equation}
\label{v4}
{\bf J}_{\omega}=-D\frac{({\bf
u}\cdot\nabla\overline{\omega}){\bf u}}{u^2}
\end{equation}
 of the vorticity current (see section \ref{sec_av}), we obtain
\begin{equation}
\label{v5}
\frac{\partial {\bf u}}{\partial t}+({\bf u}\cdot \nabla){\bf u}=-\frac{1}{\rho}\nabla p+D\frac{({\bf
u}\cdot\nabla\overline{\omega})\nabla\psi}{u^2}.
\end{equation}

{\it Remark:} With the parametrization of \ref{sec_local}, the turbulent terms in equations (\ref{v3}) and (\ref{v5}) are replaced by $D(\Delta{\bf u}-\frac{{\bf r}\cdot \nabla\overline{\omega}}{r^2}{\bf r}_{\perp})$ and $-D\frac{({\bf r}_{\perp}\cdot\nabla\overline{\omega})}{r^2}{\bf r}$ respectively.

\section{Application to the shallow-water equations}
\label{sec_sw}

Chavanis and Sommeria (2002) have developed a statistical theory of
the shallow-water (SW) equations and they have derived a set of
relaxation equations towards the statistical equilibrium state by
using a MEPP. This is a generalization of the parametrization
(\ref{mepp8}).  The relaxation equations (\ref{two1}) and (\ref{two2})
have been generalized to the SW equations by Chavanis and Dubrulle
(2006). Finally, it is straightforward to generalize the
parametrization (\ref{l5}) to the SW equations. This leads to the following set of
equations
\begin{equation}
\label{sw1}
\frac{\partial h}{\partial t}+\nabla\cdot (h{\bf u})=0,
\end{equation}
\begin{equation}
\label{sw2}
\frac{\partial {\bf u}}{\partial t}+\overline{q}h{\bf z}\times {\bf u}=-\nabla B-{\bf z}\times {\bf J}_{\omega},
\end{equation}
\begin{equation}
\label{sw3}
B=gh+\frac{{\bf u}^2}{2},\qquad \overline{q}=\frac{\omega+2\Omega}{h},
\end{equation}
\begin{equation}
\label{sw4}
\frac{\partial}{\partial t}(h \overline{q})+\nabla\cdot (h \overline{q} {\bf u})=-\nabla\cdot {\bf J}_{\omega},
\end{equation}
\begin{equation}
\label{sw5}
{\bf J}_{\omega}=-D({\bf r},t)\left (\nabla \overline{q}-\frac{{\bf u}_{\perp}\cdot \nabla \overline{q}}{u_{\perp}^2}{\bf u}_{\perp}\right ),
\end{equation}
\begin{equation}
\label{sw6}
\frac{\partial}{\partial t}(h \rho)+\nabla\cdot (h \rho {\bf u})=-\nabla\cdot {\bf J},
\end{equation}
\begin{equation}
\label{sw7}
{\bf J}=-D({\bf r},t)\left \lbrack\nabla \rho-\rho(\sigma-\overline{q})\frac{{\bf u}_{\perp}\cdot\nabla\overline{q}}{q_2 u_{\perp}^2}{\bf u}_{\perp}\right \rbrack,
\end{equation}where $q$ is the potential vorticity and $B$ is the Bernouilli function. The generalized entropies are $S[\overline{\omega},h]=-\int C(\overline{\omega})h\, d{\bf r}$. If we apply the anticipated vorticity method of section \ref{sec_av}, we get ${\bf J}_{\omega}=-(D/u^2)({\bf u}\cdot \nabla\overline{q}){\bf u}$ which is equivalent to equation (\ref{sw5}).

\section{Local conservation of angular momentum}
\label{sec_local}

The equations derived in sections \ref{sec_l} and \ref{sec_av} conserve the energy locally, but they do not conserve the angular momentum. For the sake of completeness, using similar methods, we derive here an equation that conserves the angular momentum locally. However, this equation is of little practical interest since it does not conserve the energy\footnote{It is not possible to conserve locally the angular momentum and the energy since the vorticity current ${\bf J}_{\omega}$ cannot be perpendicular to both $\nabla\psi$ and ${\bf r}$ when these two vectors are not collinear.}.

\subsection{From the MEPP}

The conservation of angular momentum $L=\int\overline{\omega}r^2\, d{\bf r}$ leads to the global constraint $\dot L=2\int {\bf J}_{\omega}\cdot {\bf r}\, d{\bf r}=0$. Applying the MEPP with the local constraint ${\bf J}_{\omega}\cdot {\bf r} =0$, and writing the variational problem in the form
\begin{eqnarray}
\label{a1}
\delta \dot S-\int \frac{\lambda({\bf r},t)}{2}\delta ({\bf J}_{\omega}\cdot {\bf r})\, d{\bf r}
-\int {\bf \zeta} ({\bf r},t)\cdot \delta \biggl (\int {\mathbf{J}}d\sigma \biggr )\, d{\mathbf{r}}\nonumber\\
-\int \frac{1}{D({\mathbf{r}},t)}\delta \biggl (\int {{\bf J}^{2}\over 2\rho} d\sigma\biggr ) \, d{\mathbf{r}}=0,
\end{eqnarray}
we obtain the optimal current
\begin{equation}
\label{a2}
{\bf J}=-D({\bf r},t)\left\lbrack \nabla\rho+\lambda({\bf r},t)\rho(\sigma-\overline{\omega}){\bf r}\right\rbrack.
\end{equation}
The Lagrange multiplier ${\zeta}$ has been eliminated, using the
condition  $\int {\bf J}\, d\sigma={\bf 0}$  of local normalization conservation.
The vorticity current is
\begin{equation}
\label{a3}
{\bf J}_{\omega}=-D({\bf r},t)\left\lbrack \nabla\overline{\omega}+\lambda({\bf r},t)\omega_2{\bf r}\right\rbrack.
\end{equation}
The evolution of the Lagrange multiplier $\lambda({\bf r},t)$ is determined
by introducing equation (\ref{a3}) in the local constraint ${\bf J}_{\omega}\cdot {\bf r} =0$. This yields ${\bf r}\cdot\nabla\overline{\omega}+\lambda\omega_2 r^2=0$ implying
\begin{equation}
\label{a6}
\lambda({\bf r},t)=-\frac{{\bf r}\cdot\nabla\overline{\omega}}{\omega_2 r^2}.
\end{equation}
Finally, we obtain the equations
\begin{equation}
\label{a7}
{\partial \rho\over \partial t}+{\bf u}\cdot\nabla\rho=\nabla \cdot \left\lbrace D({\bf r},t) \left\lbrack\nabla\rho-\rho(\sigma-\overline{\omega})\frac{{\bf r}\cdot\nabla\overline{\omega}}{\omega_2 r^2}{\bf r}\right\rbrack\right\rbrace
\end{equation}
and
\begin{equation}
\label{a8}
{\partial \overline{\omega} \over \partial t}+{\bf u}\cdot\nabla\overline{\omega}=\nabla \cdot \left\lbrace D({\bf r},t) \left \lbrack\nabla\overline{\omega}-\frac{{\bf r}\cdot\nabla\overline{\omega}}{r^2}{\bf r}\right \rbrack\right\rbrace=\frac{\partial}{\partial\theta}\left (\frac{D}{r^2}\frac{\partial\overline{\omega}}{\partial\theta}\right ).
\end{equation}
Since the diffusion coefficient is unspecified, we can take $D\propto
r^2$ to avoid dividing by zero when $r=0$. Proceeding as in Section
\ref{sec_l}, we can show that equation (\ref{a7}) conserves the
normalization, the Casimirs, the angular momentum, and increases
monotonically the mixing entropy. Furthermore, equation (\ref{a8})
dissipates all the generalized enstrophies monotonically. The
vorticity current vanishes for any axisymmetric flow
$\overline{\omega}=f(r^2)$. Starting from a non-axisymmetric initial
condition, this equation is expected to reach a steady state of the 2D
Euler equation that is axisymmetric, but its precise form cannot be
determined {\it a priori}. It depends on the dynamics and we have to solve
equation (\ref{a8}).

{\it Remark:} if we take $D=Kr^2$ and ignore the advection term, the solution of equation (\ref{a8}) is $\overline{\omega}(r,\theta,t)=\sum_n\phi_n(r)e^{in\theta}e^{-Kn^2t}$ where the $\phi_n(r)$ are determined by the initial condition $\overline{\omega}(r,\theta,0)$. We find that $\overline{\omega}(r,\theta,t)\rightarrow \phi_0(r)$ for $t\rightarrow +\infty$ where $\phi_0(r)=\frac{1}{2\pi}\int_0^{2\pi} \overline{\omega}(r,\theta,0)\, d\theta$. This is a trivial example which shows how the asymptotic state of equation (\ref{a8}) is selected.

\subsection{From the anticipated vorticity method}

We can proceed as in Section \ref{sec_av} to obtain an equation conserving locally the
angular momentum while dissipating monotonically all the generalized
enstrophies. We assume that the angular momentum is conserved
locally so that ${\bf J}_{\omega}\cdot {\bf r}=0$. This implies that
${\bf J}_{\omega}=-\lambda({\bf r},t){\bf
r}_{\perp}$ where $\lambda({\bf r},t)$ is an arbitrary function.
Substituting this relation in equation (\ref{l9}) we get $\dot S=\int
C''(\overline{\omega})\lambda({\bf r},t){\bf r}_{\perp} \cdot
\nabla\overline{\omega} \, d{\bf r}$.  If we take $\lambda({\bf
r},t)=K({\bf r},t){\bf r}_{\perp} \cdot \nabla\overline{\omega}$ with $K\ge
0$, we obtain $\dot S\ge 0$. Finally, it is relevant to write  $K=D/r^2$ where
$D({\bf r},t)\ge 0$ has the dimension of a diffusion
coefficient. Therefore ${\bf J}_{\omega}=-D({\bf
r}_{\perp}\cdot\nabla\overline{\omega}){\bf r}_{\perp}/r^2$. Substituting this expression in equation (\ref{mepp1}), we obtain
\begin{equation}
\label{a9}
{\partial \overline{\omega} \over \partial t}+{\bf  u}\cdot \nabla\overline{\omega}=\nabla\cdot \left\lbrack D \frac{({\bf r}_{\perp}\cdot \nabla\overline{\omega}){\bf r}_{\perp}}{r^2}\right\rbrack=\frac{\partial}{\partial\theta}\left (\frac{D}{r^2}\frac{\partial\overline{\omega}}{\partial\theta}\right ).
\end{equation}
This equation can also be written as
\begin{equation}
\label{a10}
{\partial \overline{\omega} \over \partial t}+{\bf  u}\cdot \nabla\overline{\omega}=\nabla\cdot \left (D \frac{{\bf r}_{\perp}\otimes {\bf r}_{\perp}}{r^2}\nabla\overline{\omega}\right ).
\end{equation}
Since $D$ is unspecified, we can take $D=K r^2$
in order to avoid dividing by zero when $r=0$. This equation conserves locally the angular momentum and decreases monotonically all the generalized enstrophies. Indeed
\begin{equation}
\label{a11}
\dot S=\int C''(\overline{\omega})\frac{D}{r^2}({\bf r}_{\perp}\cdot \nabla\overline{\omega})^2 \, d{\bf r}\ge 0.
\end{equation}
The diffusion current vanishes for any axisymmetric flow
$\overline{\omega}=f(r^2)$. Actually,  equation (\ref{a9}) is equivalent to equation
(\ref{a8}) derived from the MEPP. In 
the present case, this is obvious in view of the last equality
in equations (\ref{a8}) and (\ref{a9}) but we can also show it by making the
parallel with the calculations of Section \ref{sec_av}. Combining the identity
of vector analysis
\begin{equation}
\label{a12}
{\bf r}_{\perp}\times (\nabla\overline{\omega}\times {\bf r}_{\perp})=r^2\nabla\overline{\omega}-({\bf r}_{\perp}\cdot \nabla\overline{\omega}){\bf r}_{\perp}
\end{equation}
with the relation
\begin{equation}
\label{a13}
\nabla\overline{\omega}\times {\bf r}_{\perp}=\nabla\overline{\omega}\times ({\bf z}\times {\bf r})=(\nabla\overline{\omega}\cdot {\bf r}){\bf z}
\end{equation}
leading to
\begin{equation}
\label{a14}
{\bf r}_{\perp}\times (\nabla\overline{\omega}\times {\bf r}_{\perp})=(\nabla\overline{\omega}\cdot {\bf r}){\bf r},
\end{equation}
we find that
\begin{equation}
\label{a15}
\frac{({\bf r}_{\perp}\cdot \nabla\overline{\omega}){\bf r}_{\perp}}{r^2}=\nabla\overline{\omega}-\frac{(\nabla\overline{\omega}\cdot {\bf r}){\bf r}}{r^2}.
\end{equation}
Therefore, equation (\ref{a9}) is equivalent to equation (\ref{a8}).

\section*{References}
\begin{harvard}
\item[] Bouchet F, Venaille A 2012 Statistical mechanics of two-dimensional and geophysical flows  {\it Phys. Rep.} {\bf 515} 227-295
\item[] Chavanis P-H 2000  Quasilinear theory of the 2D Euler equation {\it Phys. Rev. Lett.} {\bf 84} 5512-5515
\item[] Chavanis P-H 2001 Kinetic theory of point vortices: Diffusion coefficient and systematic drift {\it Phys. Rev. E} {\bf 64} 026309-1-28
\item[] Chavanis P-H 2002 {Statistical mechanics of two-dimensional
vortices and stellar systems} in: ``Dynamics and Thermodynamics of Systems with
Long Range Interactions'', edited by Dauxois T, Ruffo S, Arimondo E and
Wilkens M, {\it Lect. Not. in Phys.} {\bf 602}, Springer [cond-mat/0212223]
\item[] Chavanis P-H 2003 Generalized thermodynamics and Fokker-Planck equations: Applications to stellar dynamics and two-dimensional turbulence {\it Phys. Rev. E} {\bf 68} 036108-1-20
\item[] Chavanis P-H 2006 Coarse-grained distributions and superstatistics {\it Physica A} {\bf 359} 177-212
\item[] Chavanis P-H 2009 Dynamical and thermodynamical stability of two-dimensional flows: variational principles and relaxation equations {\it Eur. Phys. J. B} {\bf 70} 73-105
\item[] Chavanis P-H, Dubrulle B 2006 Statistical mechanics of the shallow-water system with an a priori potential vorticity distribution {\it C. R. Physique} {\bf 7} 422-432
\item[] Chavanis P-H, Sommeria J 1997 Thermodynamical approach for small-scale parametrization in 2D turbulence {\it Phys. Rev. Lett.} {\bf 78} 3302-3305
\item[] Chavanis P-H, Sommeria J 1998 Classification of robust isolated vortices in two-dimensional hydrodynamics {\it J. Fluid Mech.} {\bf 356} 259-296
\item[] Chavanis P-H, Sommeria J 2002 Statistical mechanics of the shallow water system {\it Phys. Rev. E}  {\bf 65} 026302 1-13
\item[] Chavanis P-H, Sommeria J and Robert R 1996 Statistical mechanics of two-dimensional vortices and collisionless stellar systems {\it Astrophys. J.} {\bf 471} 385-399
\item[] Lynden-Bell D 1967 Statistical mechanics of violent relaxation in stellar systems {\it Month. Not. R. astron. Soc.} {\bf 136} 101-121
\item[] Miller J 1990 Statistical mechanics of Euler equations in two dimensions {\it Phys. Rev. Lett.} {\bf 65}  2137-2140
\item[] Robert R, Rosier C 1997 The modeling of small scales in two-dimensional turbulent flows: A statistical mechanics approach {\it J. Stat. Phys.} {\bf 86} 481-515
\item[] Robert R, Sommeria J 1991 Statistical equilibrium states for two-dimensional flows {\it J. Fluid Mech.} {\bf 229}  291-310
\item[] Robert R, Sommeria J 1992 Relaxation towards a statistical equilibrium state in two-dimensional perfect fluid dynamics {\it Phys. Rev. Lett.} {\bf 69} 2776-2779
\item[] Sadourny R, Basdevant C 1981 A class of operators for modelling two-dimensional turbulent diffusion {\it C. R. Acad. Sc. Paris} {\bf 292} 1061-1064
\item[] Sommeria J, Staquet C and Robert R 1991 Final equilibrium state of a two-dimensional shear layer {\it J. Fluid Mech.} {\bf 233} 661-689
\item[] Tremaine S, H\'enon M and Lynden-Bell D 1986 H-functions and mixing in violent relaxation {\it Month. Not. R. astron. Soc.} {\bf 219} 285-297
\item[] Vallis G K, Carnevale G F and Young W R 1989 Extremal energy properties and construction of stable solutions of the Euler equations {\it J. Fluid Mech.} {\bf 207} 133-152

\end{harvard}

\end{document}